\documentclass[twocolumn,
 amsmath,amssymb,
 aps, physrev,
]{revtex4-2}
\usepackage{graphicx}
\usepackage{dcolumn}% Align table columns on decimal point
\usepackage{bm}% bold math
\usepackage{color}
\usepackage{soul}
\usepackage{textcomp}
\usepackage{multirow}
\usepackage{setspace}
\usepackage[mathlines]{lineno}
\usepackage{textcomp}
\usepackage{mathcomp}
\usepackage{float}
\usepackage{longtable}
\usepackage{amssymb}% http://ctan.org/pkg/amssymb

\usepackage[T1]{fontenc}       % Use modern font encodings
\usepackage{amsmath,amsfonts}
\usepackage{epsfig}
\usepackage{rotating}
\usepackage{longtable}
\usepackage{epstopdf}
\usepackage{xcolor}
\usepackage{soul}
\usepackage{hyperref}
\hypersetup{
    colorlinks=true,       % false: boxed links; true: colored links
    linkcolor=blue,        % color of internal links
    citecolor=blue,        % color of links to bibliography
    filecolor=blue,        % color of file links
    urlcolor=blue,         % color of external links
    runcolor=blue
}

\newcommand{\ket}[1]{\left| #1 \right\rangle}

\newcommand{\average}[3]{\left\langle #1 \right| #2 \left| #3 \right\rangle}

\newcommand{\omc}{\omega_{\mathrm{c}}}

\newcommand{\UCN}{Departamento de F\'isica, Universidad Cat\'olica del Norte, Av. Angamos 0610, Antofagasta, Chile}
\newcommand{\USACH}{Departamento de F\'isica, Universidad de Santiago de Chile, Av. Victor Jara 3493, Santiago, Chile}
\newcommand{\MIRO}{Millennium Institute for Research in Optics, Concepci\'on, Chile}

\begin{document}

\title{Enhancing Infrared Laser Dissociation of Molecules with the Electromagnetic Vacuum}

\author{Johan F. Triana}
\email{johan.triana@ucn.cl}
\affiliation{\UCN}

\author{Felipe Herrera}
\email{felipe.herrera.u@usach.cl}
\affiliation{\USACH}
\affiliation{\MIRO}

\date{\today}

\begin{abstract}
Controlling bond breaking is a long-standing goal in molecular physics. Infrared nanocavities are currently being developed for reaching exotic coupling regimes of cavity QED with a few molecules, but it is not well understood how chemical reactions would proceed in such systems. 
To address this, we study infrared laser photodissociation of a single CS$_{2}$ molecule with a stretching mode that strongly interacts with a resonant infrared vacuum, subject to a strong laser field that either resonantly drives the molecule at its fundamental vibration frequency or injects photons at the cavity resonance. 
We show that the intensities required for photodissociation are significantly lower inside the cavity than in free space, with a strong dependence on the type of driving condition. 
By directly injecting photons into the cavity, the molecule dissociates with two orders of magnitude less laser energy than by directly driving the vibrational mode. 
This photodissociation enhancement is a purely quantum mechanical effect that cannot be captured semi-classically. 
The intracavity ladder climbing dynamics is substantially modified relative to free space due to vacuum-induced admixing a large number of vibrational quantum numbers and the cavity field acting as a surrogate molecular mode that strongly interacts with the dissociative vibrational motion. 
Our work provides fundamental mechanistic understanding of chemical dynamics that can be used for designing new types of nanophotonics experiments that probe single-molecule chemistry.
\end{abstract}

\maketitle 

%\section{Introduction}
%%%%%%%%%%%%%%%%%%%%%%%%%%%%%%%%%%%%%%%%%%%%%%

Controlling chemical reactions using electromagnetic fields is a long-standing goal in chemistry and physics \cite{Warren1993}.
Infrared cw lasers can be used to resonantly drive the fundamental vibration frequency to excite the lowest vibrational states of a molecule, initiating chemical processes that involve bond breaking \cite{Zhang1989,Crim1990,Elghobashi2003}. 
However, infrared photodissociation is challenging to control due to the anharmonic nature of vibrational energy levels, which tend to become far detuned from the laser frequency for highly excited levels.   
Overcoming this anharmonic blockade effect relies on multiphoton vibrational excitation using strong laser fields \cite{Duperrex1979,Bomse1979,Lima2014}, negatively chirped pulses \cite{Chelkowski1990,Lin1998,Liu1998,Marcus2006}, and optimized pulse shaping \cite{Noid1979,Matjaz1994,Korolkov1997cpl}.
In polyatomic molecules, the spectral anharmonicity can be compensated with a higher density of states at intermediate regions of the vibrational potential, which results from the interaction with other vibrational modes of the molecule \cite{Quack1978,Carmeli1980,Letokhov1981}.
The higher density of states produces a vibrational quasi-continuum that facilitates ladder-climbing to higher vibrational levels, reducing the laser intensity needed to break a molecular bond \cite{Letokhov1981}.

It is unknown how these laser chemistry concepts generalize when molecules are inside infrared cavities in the strong coupling regime of cavity QED. 
Ladder climbing processes should then occur through hybrid light-matter states, known as vibrational polaritons \cite{Long2015,Simpkins2015,George2016,Herrera2020}, which have been experimentally observed with molecular ensembles in Fabry-Perot cavities \cite{George2015,Dunkelberger2016,Xiang2019} and nanophotonic devices \cite{Metzger2019,Muller2018,Triana2022jcp,Wang2019nc,Menghrajani2019}. 
Vibrational polaritons have been associated with observed modifications of chemical reactivity in infrared cavities \cite{Thomas2019,Lather2021,Ahn2023} and plasmonic resonators \cite{Brawley2025}. 
Recently, single and few-molecule infrared nanocavities have emerged as promising light-matter interaction platforms with ultrasmall cavity mode volumes that enable strong \cite{Bell2025} and ultrastrong \cite{Yoo2021,Arul2023} coupling. %interactions with a few vibrational dipoles.
Modifications of the photodissociation process in nanocavity systems can be expected based on spectral modification of radiative heat transfer rates \cite{Suyabatmaz2025,Brawley2025}, and fundamental changes to vibrational ladder climbing due to cavity vacuum effects are expected for single molecules in ultrastrong coupling \cite{Triana2025}, but the combined influence of strong laser excitation with strong vacuum coupling effects has not been explored.

In this work, we study the influence of the infrared nanocavity vacuum on the vibrational ladder-climbing process of an individual non-polar polyatomic molecule in strong coupling.
Using carbon disulfide (CS$_{2}$) as an example \cite{Kadyan2021}, we show that vacuum-assisted infrared photodissociation leads to higher dissociation yields with respect to free space at equal driving energy. 
We demonstrate that this enhancement of photodissociation probability is a purely quantum mechanical effect by comparing with semi-classical mean-field models. 
We also show that injecting photons directly into the nanocavity field is the most energetically efficient mechanism for breaking molecular bonds with infrared laser fields resonant with the fundamental vibrational frequency.

\begin{figure}[ht]
\includegraphics[width=0.475\textwidth]{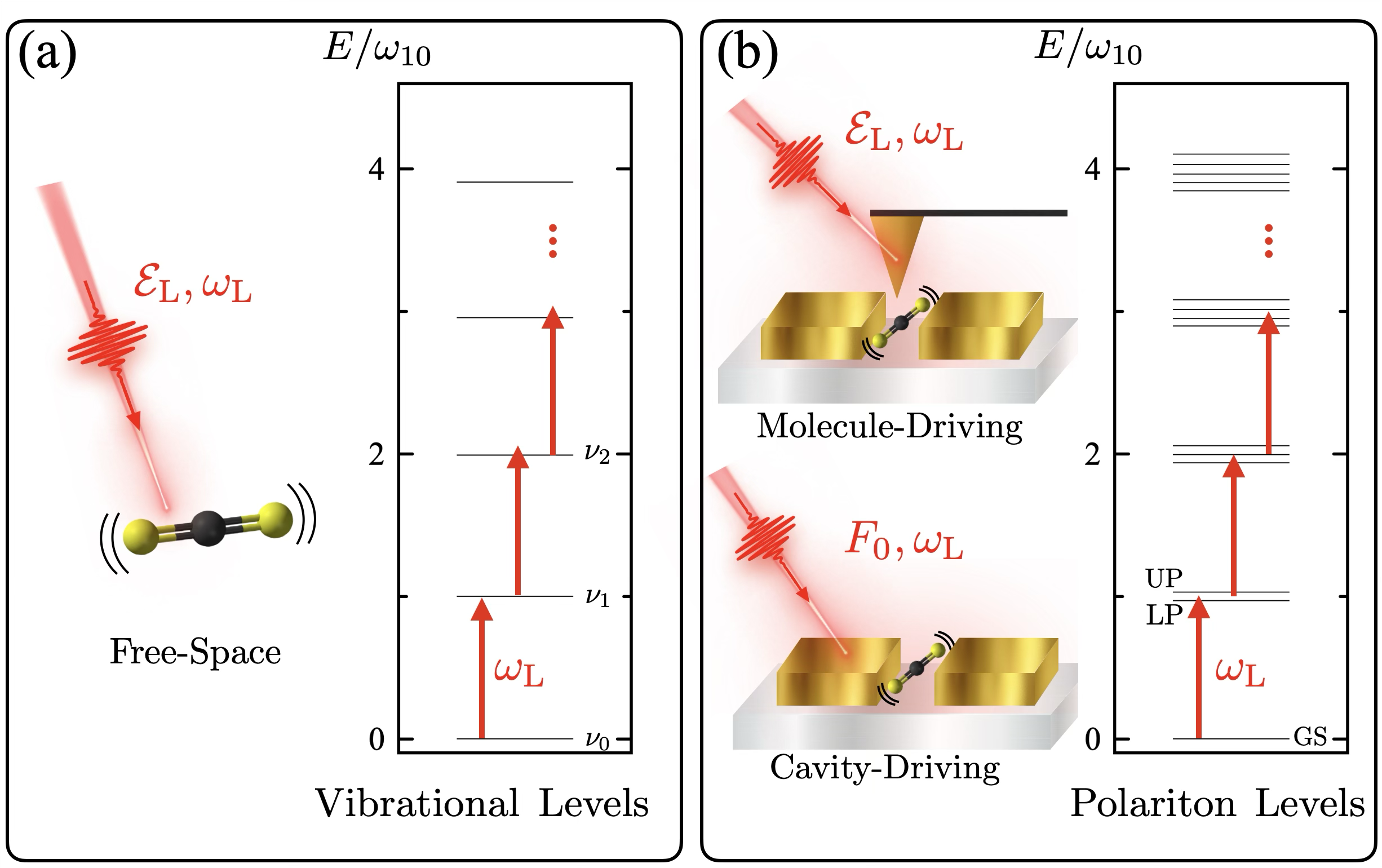} 
\caption{\textbf{Driven molecule-cavity coupled system.}
(a) CS$_{2}$ molecule in free space driven by an infrared laser. Multi-photon transitions between vibrational levels lead to dissociation. (b) CS$_{2}$ in a strongly confined infrared electromagnetic vacuum generated by metallic nanostructures. Multi-photon infrared absorption occurs between polariton levels. In the molecule-driving case, the laser drives a nanotip to excite the molecule, and in the cavity-driving scenario, the cw laser injects photons directly on the near field of the coupled system.
}
\label{fig:scheme}
\end{figure}

In free space, the simplest type of photodissociation is done with a continuous wave (cw) monochromatic laser that is resonant with the fundamental vibrational frequency  (as Fig. \ref{fig:scheme}a illustrates) \cite{Mukamel1976,Levis1999}. 
Since molecular vibrations are anharmonic, transitions between high-energy levels are suppressed, which is usually compensated by increasing the laser intensity or via frequency chirping.
Modern implementations of infrared photodissociation use a combination of strategies to optimize the dissociation yields \cite{Liu1998,Marcus2006,Noid1979,Matjaz1994,Suyabatmaz2025}.

In a cavity, the laser drives transitions between polariton eigenstates in two ways: through the electric dipole of the molecule $(\hat{d})$ or the quadrature field of the cavity ($\hat{x}$).
The dynamics of the ladder climbing process that results in bond-breaking occurs along the high density of polariton states that form a quasi-continuum, similar to the vibrational quasi-continuum established in polyatomic molecules. \cite{Quack1978,Carmeli1980,Letokhov1981}.
This high density of states enhances the ladder-climbing mechanism relative to free space, increasing dissociation yields. 
Photodissociation enhancement occurs because more states are accessible in comparison with the bare vibrational spectrum, enabling matter-matter, light-light, and light-matter transitions at infrared frequencies \cite{Herrera2017,Herrera2020}. 
These processes extend established laser-assisted chemistry concepts to cavity-QED conditions.

%%% Methods

We describe the C-S stretching mode with fundamental frequency $\omega_{10}=1514.9$ cm$^{-1}$ and dissociation energy $D_{0}\simeq28\omega_{10}$ inside an infrared nanocavity with the model Hamiltonian $\hat{\mathcal{H}}=\hat{H}_{\rm M}+\hat{H}_{\rm C}+\hat{H}_{\rm L}$, where $\hat{H}_{\mathrm{M}}=\hat{T}(\boldsymbol{q}) + \hat{V}(\boldsymbol{q})$ is the bare molecular Hamiltonian. 
$\hat{T}(\boldsymbol{q})$ is the molecular kinetic energy operator and $\hat{V}(\boldsymbol{q})$ is the potential energy with vibrational eigenstates $\ket{\nu}$, where $\nu$ is the vibrational quantum number. 
\mbox{$\hat{H}_{\rm C}=(\hat{p}^{2}+ \omega_{\mathrm{c}}^{2}\hat{x}^{2})/2 + \sqrt{2\omc}\mathcal{E}_{0}\hat{d}(q)\hat{x}$} is the bare cavity and light-matter interaction Hamiltonians, where $\hat{d}(q)$ is the dipole moment function, $\omega_{\mathrm{c}}$ is the cavity frequency and $\mathcal{E}_{0}\equiv \lambda_{g}\omc/d_{10}$, with $d_{10}=\average{\nu={1}}{\hat{d}(q)}{\nu={0}}$.
The ratio $\lambda_{g}=g/\omc$ is the conventional light-matter strength parameter \cite{Demtroder-book}, with $g=d_{10}\mathcal{E}_{0}$.

We consider two scenarios for the driving Hamiltonian $\hat{H}_{\rm L}$: (\textit{i}) the molecule is driven by the cw laser as $\hat{H}_{\rm L}^{(\mathrm{m})}=- \vec{d}(\boldsymbol{q})\cdot\mathcal{E}_{\mathrm{L}}\sin(\omega_{\mathrm{L}} t)\mathbf{\hat{e}}$, where the electric field $\vec{\mathcal{E}}_{\mathrm{L}}$ is polarized along $\mathbf{\hat{e}}$; (\textit{ii}) the cavity field is pumped by the laser as $\hat{H}^{(\mathrm{c})}_{\rm L}=F_{0}(\hat{a}^{\dagger}+\hat{a})\sin(\omega_{\mathrm{L}} t)\mathbf{\hat{e}}$, where $\hat{a}$ ($\hat{a}^{\dagger}$) is the annihilation (creation) operator of the cavity mode and  $|F_{0}|^{2}$ is proportional to the photon flux injected into the nanocavity. 
These two intracavity cases are illustrated in Fig. \ref{fig:scheme}b. 
For each case, the system is initially in the polariton ground state before the laser is turned on, and fully resonant conditions are assumed \mbox{($\omega_{\mathrm{c}}=\omega_{10}=\omega_{\mathrm{L}}$)}. 
Dissociation probabilities are computed by solving the time-dependent Schr\"odinger equation in coordinate space using the numerically exact multi-configuration time-dependent Hartree (MCTDH) method \cite{mctdhpaper,Manthe1992,Meyer1990,mctdhpack} (additional calculation details are given in Appendix \ref{app:potentials}).

%%%%%%%%%%%%%%%%%%%%%%%%
%%%%.  RESULTS
%%%%%%%%%%%%%%%%%%%%%%%%

%%%%%%%%%%%%%%%%%%%%%%%%%%%%%%%%%%%%
%%%%%%%%%%%%%%%%%%%%%%%%%%%%%%%%%%%%
\begin{figure}[t]
\includegraphics[width=0.475\textwidth]{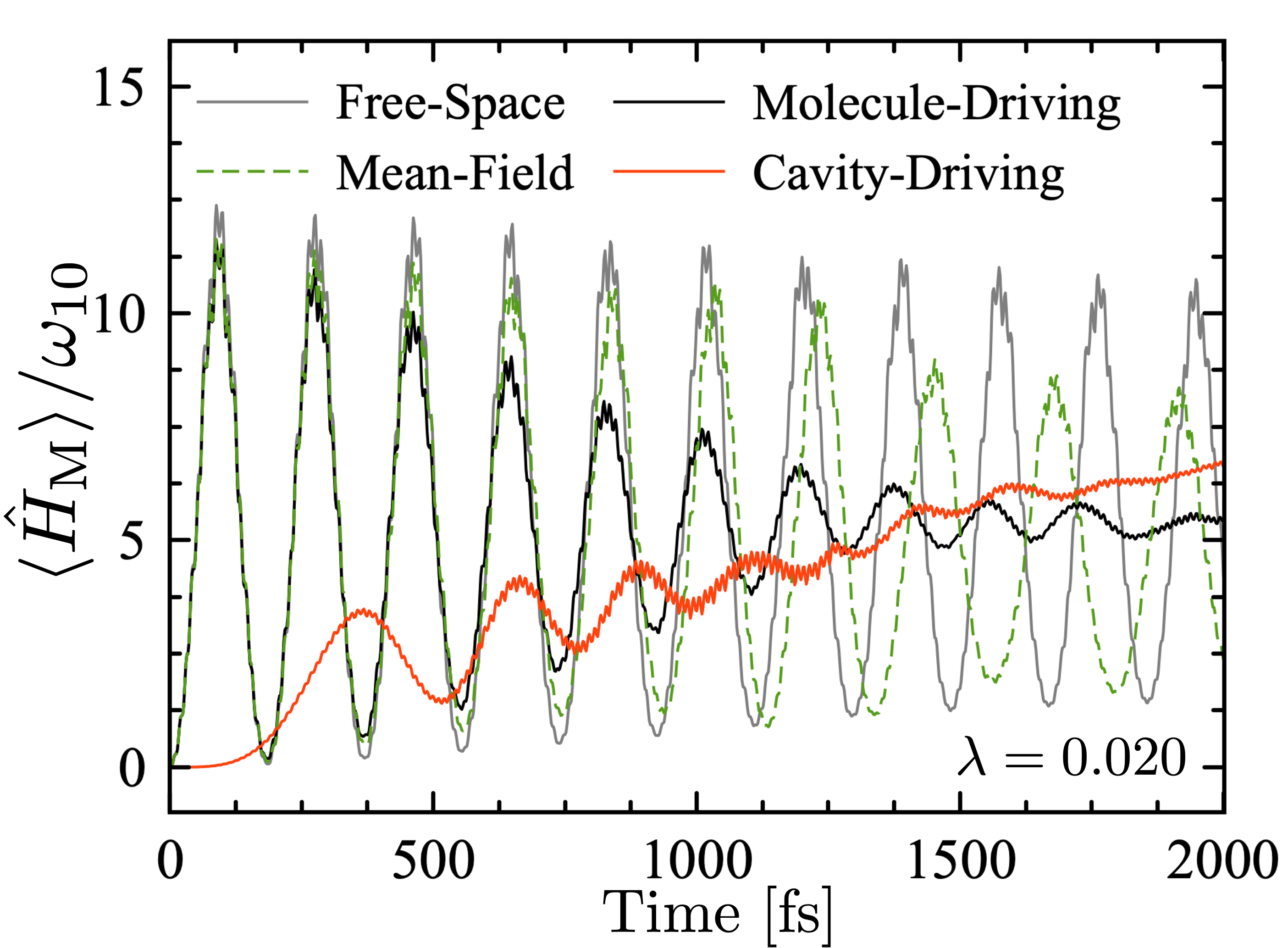}  
\caption{\textbf{Ladder-climbing dynamics.} Mean vibrational energy $\langle\hat{H}_{\mathrm{M}}\rangle$ as a function of time for free space, mean field, and intracavity scenarios. For intracavity cases, the light-matter coupling strength is $\lambda_{g}=0.02$. Molecule-driving scenarios correspond to the driving energy $E_{\mathrm{D}}\sim0.055$ aJ and for cavity-driving $E_{\mathrm{D}}\sim0.0025$ aJ. Energy in units of fundamental vibrational frequency $\omega_{10}$.
}
\label{fig:ladder}
\end{figure}
%%%%%%%%%%%%%%%%%%%%%%%%%%%%%%%%%%%
%%%%%%%%%%%%%%%%%%%%%%%%%%%%%%%%%%%

Figure \ref{fig:ladder} shows the evolution of the mean vibrational energy $\langle\hat{H}_{\mathrm{M}}\rangle$ of the CS$_{2}$ stretching mode under different driving conditions, for a cavity in strong coupling \mbox{($\lambda_{g}=0.02$)}.  
The free-space (solid gray line) and molecule-driving (solid black line) dynamics are similar and oscillate with the same frequency at short times \mbox{($t<500$ fs)}.
This oscillation is associated with the Rabi frequency of the laser driving term. 
At long times \mbox{($t>500$ fs)}, the intracavity vibrational energy stabilizes faster than in free space when the molecule is driven by the laser, because the photonic mode acts as an effective reservoir for vibrational energy. 
This type of energy exchange does not occur in free space. 
For the cavity-driving scenario, the energy also oscillates but has an overall growth with time that does not stabilize under cw driving, as expected for a driven harmonic oscillator \cite{Hanggi2018,Lopez2018}.
The energy needed to reach the same value of $\langle\hat{H}_{\mathrm{M}}\rangle$ is two orders of magnitude smaller when the cavity field is driven, relative to the molecule-driving scenario. 
This suggests that equivalent bond-breaking probabilities can be achieved by injecting photons into the cavity with much lower intensities than by directly pumping the molecule.

To assess whether these results can be explained classically, we define the effective mean-field light-matter interaction term $\hat{H}^{(\mathrm{MF})}_{\mathrm{L}}=\hat{H}_{\mathrm{L}}^{(\mathrm{m})}+\sqrt{2\omc}\mathcal{E}_{0}\hat{d}(q)\langle\hat{x}(t)\rangle_{\mathrm{m}}$, which describes the molecule driven by the laser and the classical cavity quadrature $\langle\hat{x}(t)\rangle_{\mathrm{m}}$. 
The expectation value $\langle\hat{x}(t)\rangle_{\mathrm{m}}$ is computed from the total polariton wave function. 
The mean field evolution of $\langle\hat{H}_{\mathrm{M}}\rangle$ (dashed green line in Fig. \ref{fig:ladder}) is similar to the free-space evolution, because an effective laser cannot describe the energy redistribution to photonic degrees of freedom, suppressing the fast stabilization observed when the cavity field is fully quantum.   
This is evidence that vibrational ladder climbing is strongly influenced by quantum effects of the cavity. 
Additional mean-field results are given in Appendix \ref{app:meanfield}.

Figure \ref{fig:photodissociation} shows the long-time ($t=1.5$ ps) intracavity photodissociation probability $P_{\mathrm{diss}}$ of CS$_{2}$ as a function of driving energy $E_{\mathrm{D}}$ and coupling strength $\lambda_{g}$. 
The driving energy is computed as $E_{\mathrm{D}}=\sqrt{\langle\hat{H}_{\mathrm{L}}^{2}\rangle}$ both in the molecule and cavity-driving scenarios (see additional details in Appendix \ref{app:drivingenergy}). 
The behavior of $P_{\rm diss}$ is smooth across the parameter space but a dissociation threshold can be identified as a function of driving energy for a given value of $\lambda_{g}$. 
For molecule driving (Fig. \ref{fig:photodissociation}a), there is a well-defined $\lambda_{g}\to0$ (free space) threshold at $E_{\rm D}\approx0.06$ aJ, which corresponds to an equivalent laser intensity of $\sim10^{13}$ W/cm$^{2}$. 
As the coupling strength $\lambda_{g}$ increases, this threshold intensity is reduced by an order of magnitude. 
Far beyond the threshold, $P_{\rm diss}$ becomes independent of $\lambda_{g}$.
In this regime, the intensity is strong enough to overcome anharmonic blockade effects. 
In the cavity-driving scenario (Fig. \ref{fig:photodissociation}b), there is also a well-defined dissociation threshold with respect to the driving energy at $\lambda_{g}=0.005$ ($E_{\rm D}\approx0.003$ aJ, equivalent to a laser intensity of $\sim10^{10}$ W/cm$^{2}$). 
The driving energy threshold is decreased rapidly as $\lambda_{g}$ increases. 
In general, the thresholds are at least an order of magnitude lower than the molecule-driving case, which can be equivalent to a reduction of up to three orders of magnitude in laser intensity. 
These results show that ladder climbing along vibrational polariton levels is qualitatively different from conventional vibrational ladder climbing in free space.

%%%%%%%%%%%%%%%%%%%%%%%%%%%%%%%%%%%%
%%%%%%%%%%%%%%%%%%%%%%%%%%%%%%%%%%%%
%\onecolumngrid
\begin{figure}[t]
\includegraphics[width=0.475\textwidth]{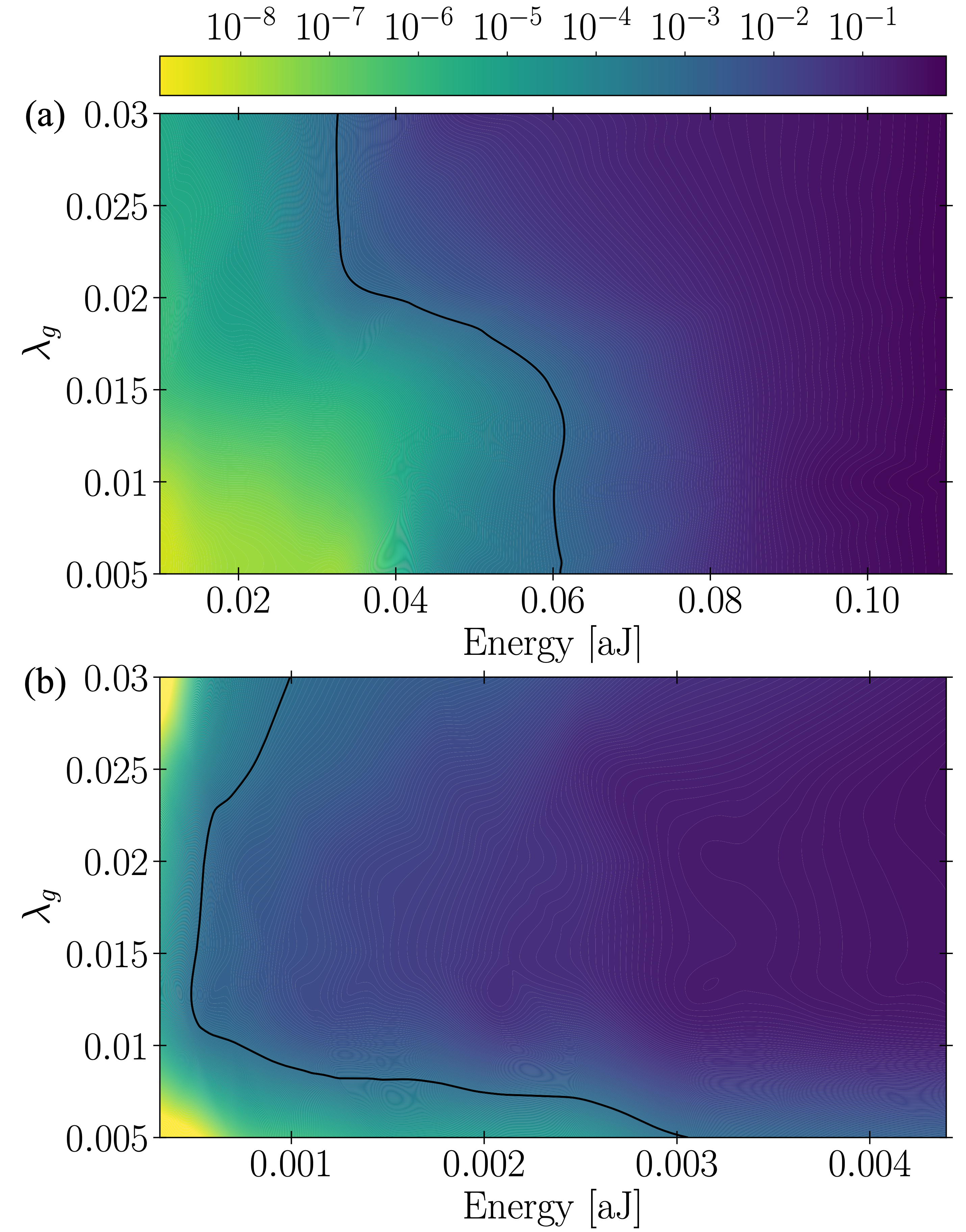}  
\caption{
\textbf{Dissociation probability maps.} (a) $P_{\mathrm{diss}}$ as a function of driving energy and coupling strength $\lambda_g$ for the molecule-driving scenario at $t=1.5$ ps; (b) same as in panel (a) for the cavity-driving scenario. Dissociaiton threshold contours for $P_{\mathrm{diss}}^{\rm (th)}=10^{-3}$ are shown (black solid lines). Energy in attojoules (aJ). For reference $0.01$ aJ $\equiv 23$ THz.
}
\label{fig:photodissociation}
\end{figure}
%%%%%%%%%%%%%%%%%%%%%%%%%%%%%%%%%%%
%%%%%%%%%%%%%%%%%%%%%%%%%%%%%%%%%%%

\begin{figure}[t]
\includegraphics[width=0.485\textwidth]{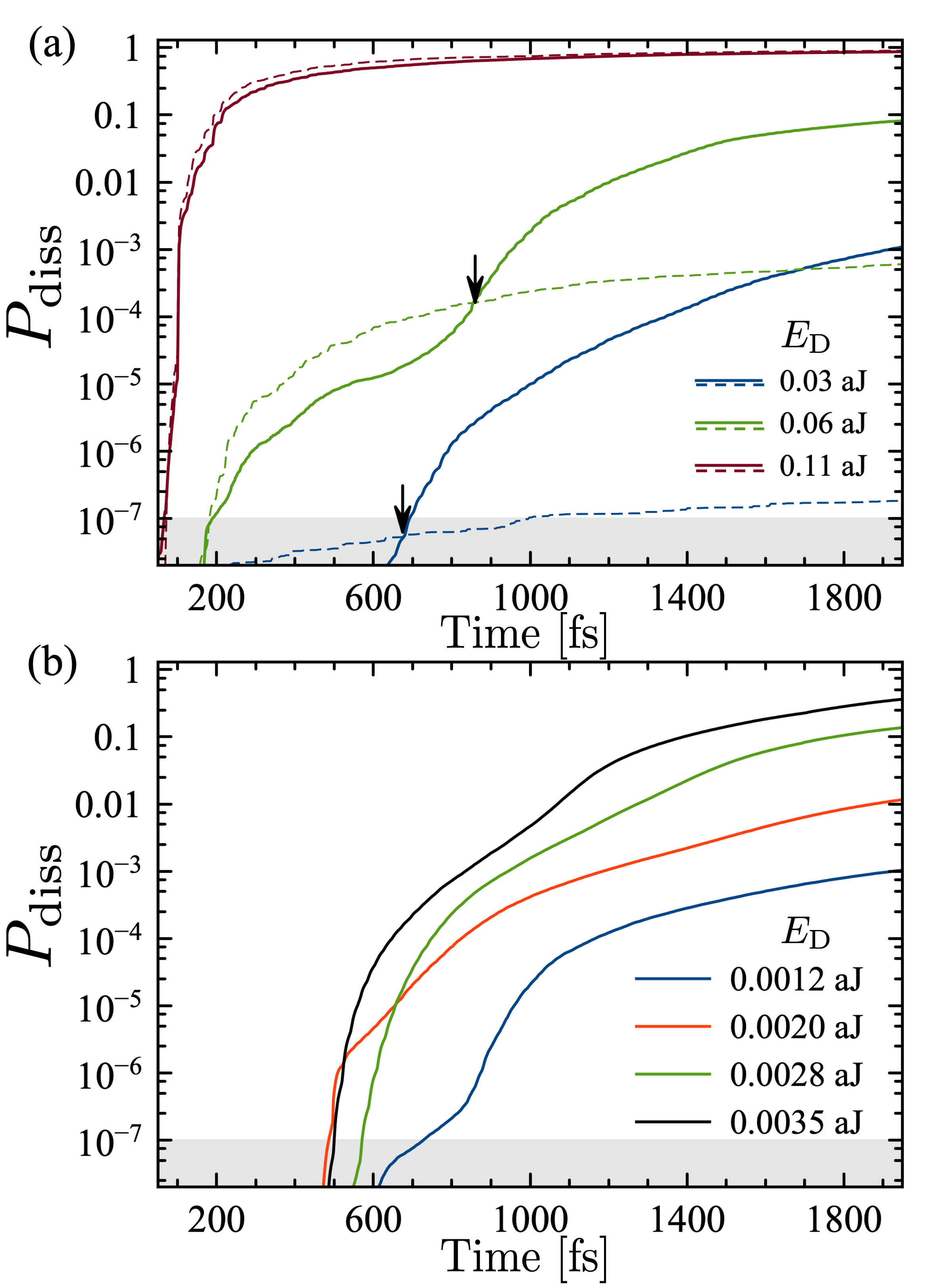}  
\caption{\textbf{Time-dependent dissociation probability.} Photodissociation probability for different driving energies $E_{\mathrm{D}}$ in aJ for (a) the molecule-driving and (b) the cavity-driving scenarios. Dashed lines for the corresponding color of laser intensity represent the free-space scenario. In both panels $\lambda_g=0.025$. 
The shaded region corresponds to the numerical error threshold.  
}
\label{fig:dissociation}
\end{figure}

Figure \ref{fig:dissociation}a shows the evolution of $P_{\mathrm{diss}}$ for an intracavity molecule directly driven with different strengths $E_{\mathrm{D}}$ (solid lines). 
The system is in strong coupling with \mbox{$\lambda_g=0.025$}. 
For comparison, results for free space at the same driving energies are shown (dashed lines). 
For $E_{\mathrm{D}}=0.03$ aJ, the system is around the dissociation threshold in Fig. \ref{fig:photodissociation}a. In comparsion with free space, $P_{\mathrm{diss}}$ increases more slowly than free space initially, but after a few hundred femtoseconds it becomes higher than free space, reaching values 10$^{4}$ times higher at long times ($t>1.5$ ps). 
Doubling the driving energy ($E_{\mathrm{D}}=0.06$ aJ) gives qualitatively similar comparisons between intracavity and free space, but there are still significant long-time differences in $P_{\mathrm{diss}}$. 
The slow increase in photodissociation probability before the activation time (black arrows) results from the light-matter interaction strength, and becomes faster for larger $\lambda_{g}$ (see additional details in Appendix \ref{app:blockade}).
For high enough driving energy, there are essentially no differences between the intracavity and free space photodissociation dynamics ($E_{\mathrm{D}}=0.11$ aJ). 
We attribute this behavior to the laser intensity being strong enough to overcome anharmonic blockade effects via multiphoton excitations.

Figure \ref{fig:dissociation}b shows the evolution of $P_{\mathrm{diss}}$ for the cavity-driving scenario in strong coupling (\mbox{$\lambda_{g}=0.025$}) subject to different driving energies $E_{\mathrm{D}}$ above threshold.
As expected, bond-breaking probabilities increase as the driving energy grows and more photons are injected into the nanocavity. 
Similar to the molecule-driving scenario, $P_{\mathrm{diss}}$ initially increases slowly due to the light-matter interaction timescale dictated by the Rabi splitting. 
This reveals how effectively the infrared vacuum enhances the vibrational ladder-climbing mechanism in a nanocavity, even with driving energies much lower than in the molecule-driving case.

%%%%%%%%%%%%%%%%%%%%%%%%%%%%%%%%%%%%
%%%%%%%%%%%%%%%%%%%%%%%%%%%%%%%%%%%%
%%%%%%%%%%%%%%%%%%%%%%%%%%%%%%%%%%%
%%%%%%%%%%%%%%%%%%%%%%%%%%%%%%%%%%%

% Summary and conclusions. 
In summary, we showed that infrared laser photodissociation of single molecules has fundamentally different dynamics inside a cavity than in free space, due to vacuum-assisted ladder-climbing transitions that occur in vibrational strong coupling. 
The details of the photodissociation process, such as the activation time and dissociation yield, depend on whether the laser drives directly the vibrational degrees of freedom or the cavity quadrature. In both cases, less laser energy is required to achieve a target dissociation outcome than in free space.  
This behavior resembles the trends found when comparing free-space infrared photodissociation of diatomic and polyatomic molecules, where the higher density of states and couplings of polyatomic molecules facilitate vibrational ladder-climbing \cite{Quack1978,Carmeli1980,Letokhov1981}. 
In our picture, the photonic mode can be seen as an auxiliary mode that facilitates the activation of the molecular mode. 
This vacuum-assisted molecular activation is a purely quantum mechanical effect that cannot be captured with a semi-classical mean-field treatment of the problem (see details in Appendix \ref{app:meanfield}).

The treatment used here is based on the multipolar formulation of cavity QED in strong coupling \cite{Hernandez2019,Triana2020,Triana2025}, which for this photodissociation problem gives equivalent results to an alternative treatment based on the Pauli-Fierz model \cite{Flick2017,Ruggenthaler2023}.  
Specific model comparisons are given in Appendix \ref{app:paulifierz}.
Theory improvements would require an accurate treatment of the photonic modes of nanocavities, including the effects of photon loss and dispersion within a macroscopic QED formulation \cite{Candelas2024,Hsu2025,Rema2025}. 
Simpler Lindblad quantum master equations in coordinate space could also be formulated to describe dissociation in multimode dissipative cavities using available numerical methodologies \cite{Triana2022mc}, which is subject for future work.

Infrared cavity photodissociation can be potentially simpler to implement in experiments compared to UV photodissociation because it involves only the ground electronic state, which has a well-known dissociation energy. 
Currently, infrared nanocavities are being developed for reaching exotic regimes of light-matter interaction, such as ultrastrong coupling with a few vibrational dipoles \cite{Yoo2021}.  
Reaching strong coupling at the single-molecule level at mid-infrared wavelengths has yet to be demonstrated, and our results demonstrate that photochemistry can be fundamentally different in this limit.
% which allows us to predict future measurements in the mid-infrared regime at the single-molecule level.  
%
Our work can significantly contribute to the design of new chemical control mechanisms that exploit purely quantum effects to reduce energy consumption in photodissociation, as well as to the understanding of how photonic degrees of freedom influence the barrier crossing dynamics in molecular systems.

%\begin{acknowledgements}
%\textit{Acknowledgements}$-$ 
We acknowledge fruitful discussions with Johannes Schachenmayer and Blake Simpkins. 
J.F.T. is supported by ANID-Fondecyt Iniciaci\'on 11230679 and ``N\'ucleo de Investigaci\'on No.7 UCN-VRIDT 076/2020, N\'ucleo de modelaci\'on y simulaci\'on cient\'ifica (NMSC)''. 
F.H. is supported by ANID through grants Fondecyt Regular 1221420 and Millennium Science Initiative Program ICN17\_012. 
%\end{acknowledgements}

\bibliographystyle{apsrev4-1}
\bibliography{biblio}

\clearpage
%\begin{widetext}
\appendix
%\input{suppmattext.tex}

%%%%%%%%%%%%%%%%%%%%%%%%%%%%%%%%%%%%%%%%%%%%%%%%%%%%%%%%%%%%%%%%%%%%%%
\section{Ab-Initio Modeling of CS$_2$ $\tilde{\rm X}^{1}\Sigma_{\rm g}^{+}$ Ground State Intracavity Dynamics}
\label{app:potentials}

\begin{figure}[t]
\includegraphics[width=0.475\textwidth]{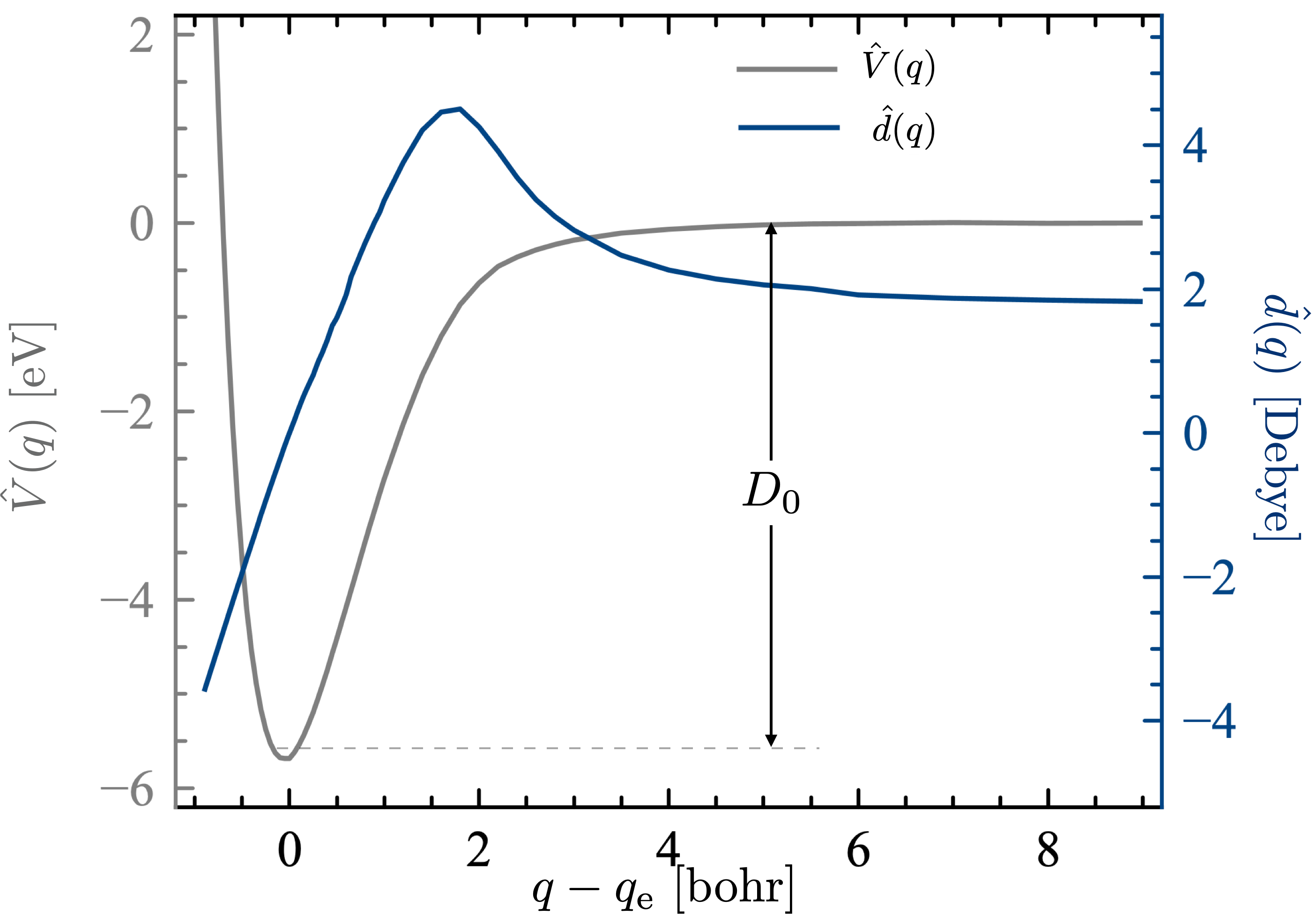}
\caption{\textit{ab initio} potential energy curve (gray) and electric dipole moment function (blue) of the ground electronic state $\tilde{\rm X}^{1}\Sigma_{\rm g}^{+}$ for CS stretching mode of CS$_{2}$ molecule as a function of bond distance coordinate $q$ respect to equilibrium $q_{\mathrm{e}}$.
$D_0$ corresponds to the dissociation energy.
}
\label{fig:cs2 potential}
\end{figure}

The potential energy curve $\hat{V}(q)$ of the electronic ground state of CS$_{2}$ molecule as a function of C-S stretching mode and its corresponding dipole moment function $\hat{d}(q)$ were calculated using the electronic structure package MOLPRO \cite{MOLPRO} using the multireference configuration interaction (MRCI) method with the aug-cc-pVQZ basis set, considering the CS$_{2}$ molecule aligned along $z$ axis, which are shown in Fig. \ref{fig:cs2 potential}.

For quantum dynamics calculations, we implement the multi-configuration time-dependent Hartree (MCTDH) method \cite{mctdhpaper,mctdhpack} to solve the time-dependent Schr\"odinger equation in the coordinate representation, using the light-matter Hamiltonian $\hat{\mathcal{H}}$.
The coordinate $\hat{q}$ is defined in a sin-DVR primitive basis with $N_{q}=496$ grid points for bond distances between $q=2.1$ a.u. and $q=12.0$ bohr. The bare photonic mode $\hat{x}$ is described by a single harmonic oscillator and is defined in a HO-DVR primitive basis with $N_x = 1050$ grid points between $x = -350.0$ and $x = 350.0$. The number of single-particle functions required to converge numerical calculations depends on the light-matter coupling and electric field amplitude. According to these parameters, $n_{i}$ was chosen between $20$ and $80$ for each coordinate. A complex absorbing potential (CAP) is included at $q=8.0$ bohr to avoid unphysical reflections.

%%%%%%%%%%%%%%%%%%%%%%%%%%%%%%%%%%%%%%%%%%%%%%%%%%%%%%%%%%%%%%%%%%%%%%
\section{Quantum dynamics simulations in the Mean-Field approximation}
\label{app:meanfield}

We perform a mean-field analysis to understand how the quantum effects of the photonic mode influence the modifications of the infrared photodissociation probability. 
The mean-field Hamiltonian associated with molecule-driving scenario $\hat{\mathcal{H}}_{\mathrm{MF}}^{\mathrm{(mol)}}$ is given by
\begin{equation}
\hat{\mathcal{H}}_{\mathrm{MF}}^{(\mathrm{mol})} = \hat H_{\rm M} + \hat{H}_{\rm L} + \sqrt{2\omc}\mathcal{E}_{0}\hat{d}(q)\langle\hat{x}(t)\rangle_{\mathrm{M}},
\end{equation}
where $\langle\hat{x}(t)\rangle_{\mathrm{M}}$ is the expectation value of coordinate $x$ for the photonic mode, which is calculated with the total wave function that results from the time-dependent Schr\"odinger equation solution with the light-matter Hamiltonian $\hat{\mathcal{H}}$. 
The mean-field approximation can be seen as a modulation of the cw laser, i.e., the molecule is driven by an effective laser that results in an effective driven Hamiltonian
$$
\hat{H}_{\mathrm{L-M}}^{\mathrm{(eff)}} = -\hat{d}(q)\left[ \hat{\mathcal{E}}_{\rm L}\sin(\omega_{\mathrm{c}}t)  - \sqrt{2\omc}\mathcal{E}_{0}\langle\hat{x}(t)\rangle_{\mathrm{M}}\right].
$$
Similarly, the mean-field Hamiltonian for the cavity-driving scenario reads 
\begin{equation}
\hat{\mathcal{H}}_{\mathrm{MF}}^{(\mathrm{cav})} = \hat H_{\rm M} + \sqrt{2\omc}\mathcal{E}_{0}\hat{d}(q)\langle\hat{x}(t)\rangle_{\mathrm{C}},
\label{eq:mfcavity}
\end{equation}
where 
$
\langle\hat{x}(t)\rangle_{\mathrm{C}}=\sqrt{2\omega_{\rm c}} F_{0} \left[ t \sin(\omega_{\rm c}t) \right] % -  \sin(\omega_{\rm c}t)/\omega_{\rm c}\right]
$
is the solution of the driven cavity Hamiltonian
\begin{equation}
\hat{H}_{\rm CL}=-\frac{1}{2}\frac{\partial^{2}}{\partial \hat{x}^{2}} + \frac{1}{2}\omega_{\rm c}\hat{x}^{2} + \sqrt{2\omega_{\rm c}}F_{0}\hat{x}\sin(\omc t), 
\label{eq:mfcavity2}
\end{equation}
and the effective laser Hamiltonian in this scenario is given by 
$$
\hat{H}_{\mathrm{L-C}}^{\mathrm{(eff)}} = \sqrt{2\omc}\mathcal{E}_{0}\hat{d}(q)\langle\hat{x}(t)\rangle_{\mathrm{C}}.
$$

\begin{figure*}[th]
\centering
\includegraphics[width=0.85\textwidth]{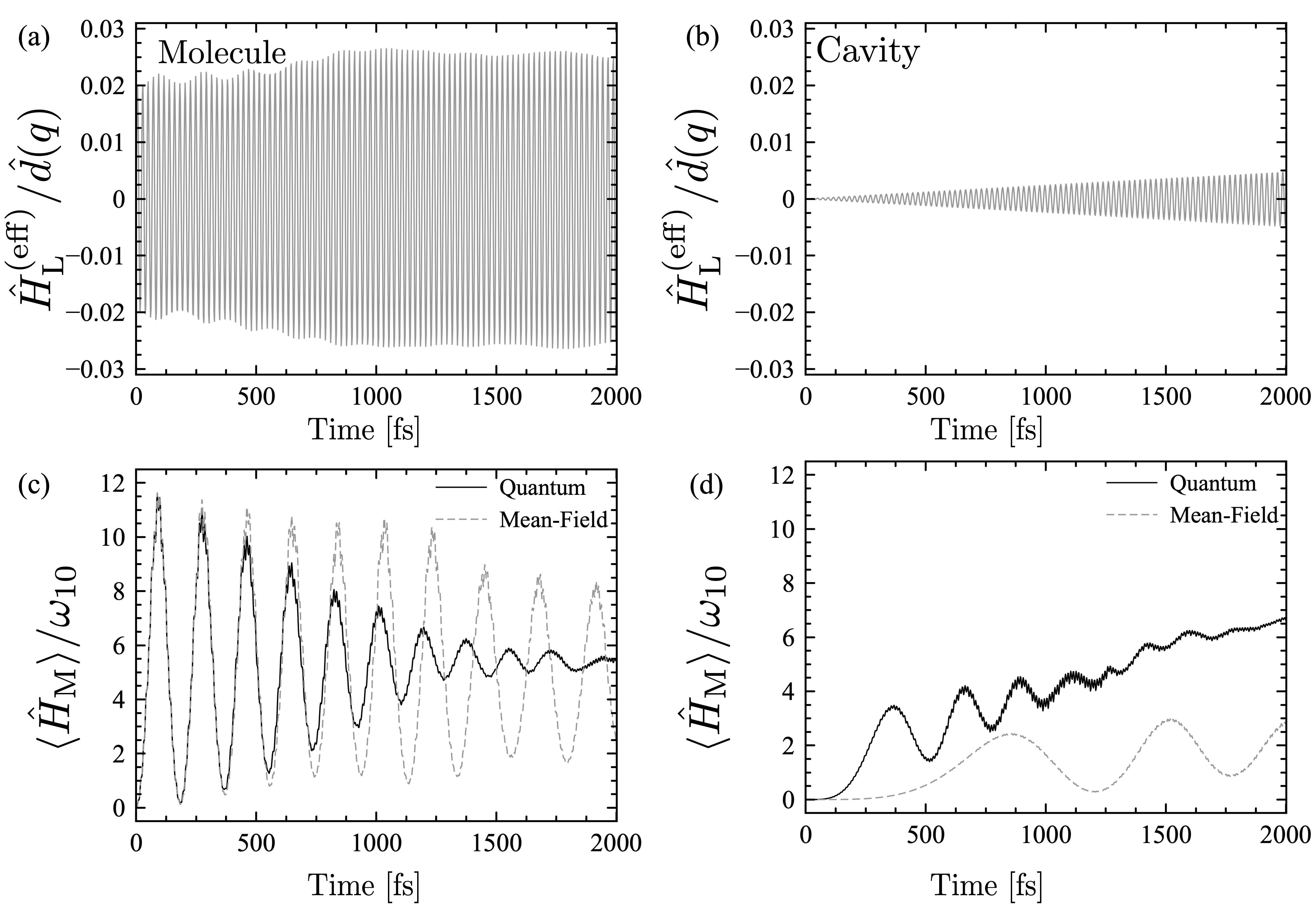}
\caption{\textbf{Mean-field approximation.} Expectation value of the normalized effective driven Hamiltonian $\hat{H}_{\mathrm{L-X}}^{\mathrm{(eff)}}/\hat{d}(q)$ as a function of time for (a) molecule-driving and (b) cavity-driving scenarios, with X=M, C. Normalized vibrational mean energy as a function of time for (c) molecule-driving and (d) cavity-driving scenarios. In all cases $\lambda_{g}=0.02$ and molecule-driving scenario correspond to $E_{\mathrm{D}}\sim0.055$ aJ and for cavity-driving $E_{\mathrm{D}}\sim0.0025$ aJ. 
}
\label{fig:mfmolecule} 
\end{figure*}

Figures \ref{fig:mfmolecule}a-\ref{fig:mfmolecule}b show the expectation values of the effective driven Hamiltonian for the molecule and cavity driving scenario, respectively. 
The most notable difference refers to the laser field strength in the cavity-driving scenario, which, on average, is approximately one order of magnitude lower than in the molecule-driving case. 
The functional shape of $\hat{H}_{\mathrm{L-M}}^{\mathrm{(eff)}}$ in Fig. \ref{fig:mfmolecule}a varies along time, but its amplitude does not increase continuously, in comparison with the cavity-driving case that depends linearly on time. 

Figure \ref{fig:mfmolecule}c shows the normalized molecular mean energy $\langle\hat{H}_{\mathrm{M}}\rangle/\omega_{10}$ for the mean-field (dashed line) and quantum (solid line) scenarios of the molecule-driving scenario. 
The mean molecular energy in the mean-field approximation tends to stabilize more slowly than the intracavity scenario, similar to the free-space scenario described in the main text. 
However, the molecular mean energy in the mean-field cavity-driving case increases as the effective driving energy increases at a small rate in comparison with the complete quantum scenario, as Fig. \ref{fig:mfmolecule}d shows. 
The latter is because the number of photons injected into the cavity interacts more efficiently than by driving the molecule directly.

%%%%%%%%%%%%%%%%%%%%%%%%%%
%%%%%%%%%%%%%%%%%%%%%%%%%%
%%%%%%%%%%%%%%%%%%%%%%%%%%
\section{Driving energy}
\label{app:drivingenergy}

In the molecule-driving scenario, the driving energy $E_{\rm D}$  of the driving Hamiltonian \mbox{$\hat{H}_{\rm L}=- \vec{d}(\boldsymbol{q})\cdot\mathcal{E}_{\mathrm{L}}\sin(\omega_{\mathrm{L}} t)\mathbf{\hat{e}}$} reads
\begin{equation}
E_{\rm D} = \sqrt{\langle\hat{H}^{2}_{\mathrm{L}}\rangle} = \mathcal{E}_{\mathrm{L}} \sqrt{\langle d^{2}(\boldsymbol{q})\sin^{2}(\omega_{\mathrm{L}} t) \rangle} \approx \sqrt{\frac{ \langle d^{2}(\boldsymbol{q})\rangle}{2}}\mathcal{E}_{\mathrm{L}},
%
%\frac{\mathcal{E}_{\mathrm{L}}}{q_{2}-q_{1}}\cdot\int_{q_{1}}^{q_{2}}d(q)\,\mathrm{d}q \approx 0.85\mathcal{E}_{\mathrm{L}},
\end{equation}
where
\begin{equation}
\langle d^{2}(\boldsymbol{q})\rangle=\frac{1}{q_{2}-q_{1}}\int_{q_{1}}^{q_{2}}d^{2}(\boldsymbol{q})\,\mathrm{d}\boldsymbol{q} \approx 0.85,
\end{equation}
with $q_{1}=2.1$ bohr, the first point of the molecular grid, and $q_{2}=8.0$ bohr corresponds to the CAP position.  
Thus, the driving energy as a function of electric field amplitude in the molecule-driving scenario is \mbox{$E_{\rm D} \simeq 0.65\mathcal{E}_{\mathrm{L}}$}.

In the cavity-driving scenario, the driving Hamiltonian reads
$$
\mbox{$\hat{H}_{\rm L}=F_{0}(\hat{a}^{\dagger}+\hat{a})\sin(\omega_{\mathrm{L}} t)=\sqrt{2\omega_{\mathrm{c}}} F_{0} \hat{x}\sin(\omega_{\mathrm{L}} t)$,}
$$
and the corresponding driving energy $E_{\mathrm{D}}$ is the equivalent energy of a laser-driven empty cavity 
\begin{equation}
E_{\rm D} = \sqrt{\langle\hat{H}^{2}_{\mathrm{L}}\rangle} = F_{0}\sqrt{2\omega_{\mathrm{c}} \langle \hat{x}^{2}\sin^{2}(\omega_{\mathrm{L}} t) \rangle} \approx  \sqrt{\omega_{\mathrm{c}}\overline{\langle \hat{x}^{2} \rangle}} F_{0} ,
\end{equation}
where $\overline{\langle \hat{x}^{2} \rangle}$ is the time-averaged expectation value of $\hat{x}^{2}$ given by  
\begin{widetext}
\begin{equation}
\overline{\langle \hat{x}^{2} \rangle} = \frac{1}{2\omega_{\mathrm{c}}t_{f}}\frac{F_{0}^{2}}{\omega_{\rm c}^{2}}\int_{0}^{t_{f}}\left[ \omega_{\rm c}^{2}t^{2}\cos^{2}(\omega_{\rm c}t) - \omega_{\rm c}t\sin(2\omega_{\rm c}t) + \sin^{2}(2\omega_{\rm c}t) +\frac{\omega_{\rm c}^{2}}{F_{0}^{2}} \right]\mathrm{d}t, 
\end{equation}
\end{widetext}
which is obtained by solving the time-dependent Schr\"odinger equation of a driven empty cavity with Hamiltonian in Eq. (\ref{eq:mfcavity2}).
Hence, the mean value $\overline{\langle \hat{x}^{2} \rangle}$ as a function of external driving energy $F_{0}$ and cavity frequency $\omega_{\rm c}$ in the fully resonant scenario can be expressed as
\begin{align}
\overline{\langle \hat{x}^{2} \rangle} &\simeq \frac{1}{2\omega_{\mathrm{c}}} \left\lbrace
1 +
 \left[ \frac{F_{0}^{2}}{t_{f}}\int_{0}^{t_{f}}t^{2}\cos^{2}(\omega_{\rm c}t) \,\mathrm{d}t  \right] \right\rbrace
\end{align}
which we compute for $t_{f}=1.5$ ps, the time at which dissociation probabilities are shown in Fig. \ref{fig:photodissociation}.

%%%%%%%%%%%%%%%%%%%%%%%%%%%%
%%%%%%%%%%%%%%%%%%%%%%%%%%%%
%%%%%%%%%%%%%%%%%%%%%%%%%%%%

\section{Dissociation Probabilities as a function of $\lambda_{g}$}
\label{app:blockade}

\begin{figure}[t]
\centering
\includegraphics[width=0.5\textwidth]{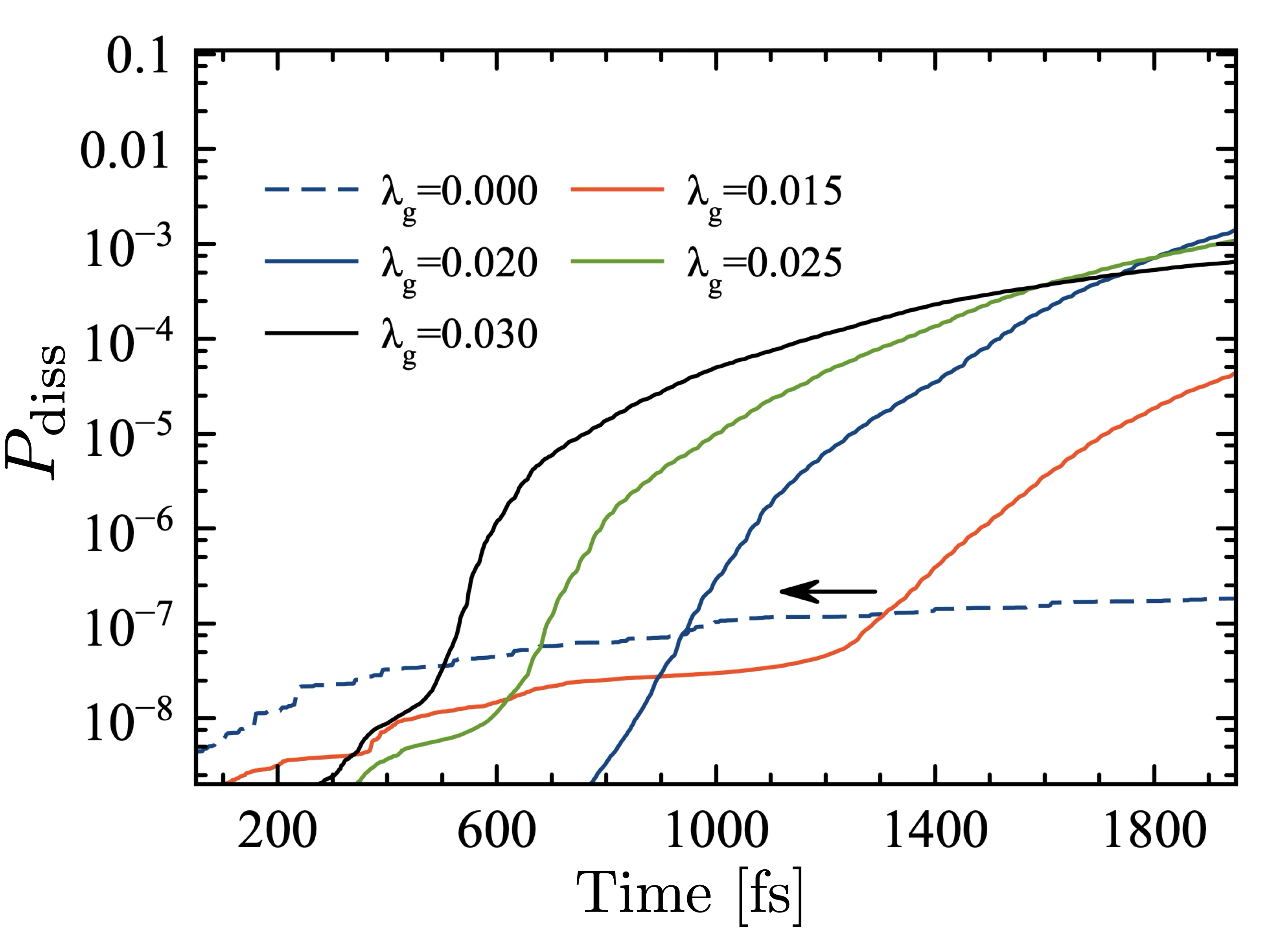}
\caption{\textbf{Time-dependent dissociation probability.} 
Intracavity dissociation probabilities for $E_{\mathrm{D}}=0.03$ aJ and different light-matter couplings $\lambda_g$. The arrow indicates the shift in the activation time for the intracavity photodissociation probability as $\lambda_{g}$ increases.}
\label{fig:blockade} 
\end{figure}

Figure \ref{fig:blockade} shows $P_{\mathrm{diss}}$ for different coupling strengths and a driving energy $E_{\mathrm{D}}=0.03$ aJ.
In free space (dashed line), the dissociation probability is very low and remains almost constant around $10^{-6}$ over the entire propagation time. 
In contrast, intracavity dissociation probabilities (solid lines) increase by several orders of magnitude and exhibit a remarkable monotonic coupling-dependent trend, as illustrated by the arrow. 
This monotonic behavior reveals that the time at which $P_{\mathrm{diss}}$ begins to rise with respect to free space, which we call activation time, is determined by the value of $\lambda_{g}$.
The latter implies that intracavity photodissociation processes are governed by light-matter interactions instead of initial polariton blockade effects (see Fig. \ref{fig:scheme}b), which can occur in the weak-field limit.

\mbox{} 

%\mbox{}  \\ 

%\clearpage
%%%%%%%%%%%%%%%%%%%%%%%%%%%%%%%%%%%%%%%%%%%%%%%%%%%%%%%%%%%%%%%%%%%%%%
\section{Comparison with Pauli-Fierz model}
\label{app:paulifierz}

\begin{figure}[b]
\centering
\includegraphics[width=0.5\textwidth]{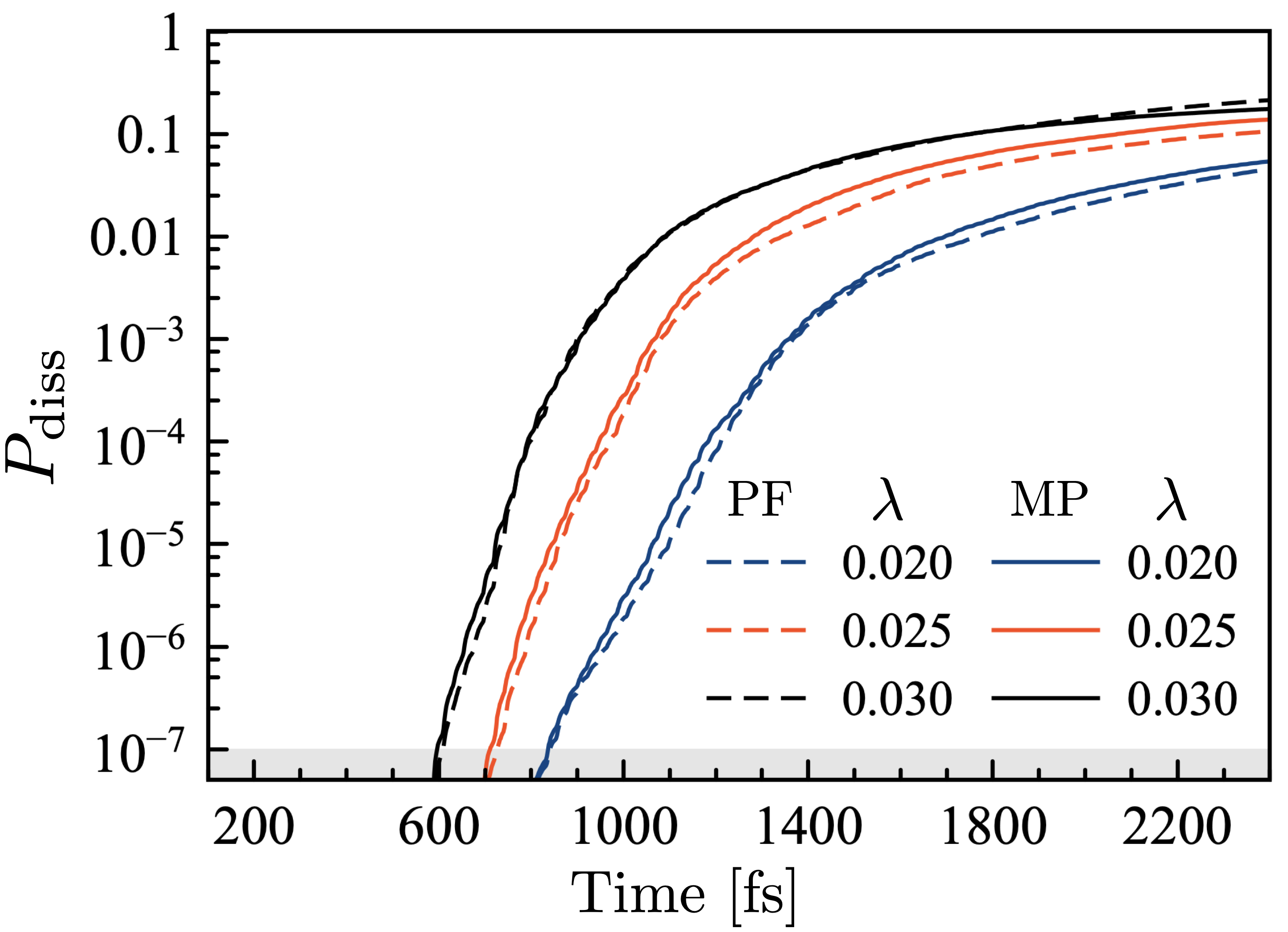}
\caption{\textbf{Pauli-Fierz vs. Multipolar QED dissociation probability.} Dissociation probability for different values of $\lambda_{g}$ with $I_{0} = 8.0\times10^{12}$ W/cm$^{2}$  ($E_{\mathrm{D}}=0.055$ aJ), for the molecule-driving scenario. Solid lines represent $P_{\mathrm{diss}}$ calculated without DSE, and dashed lines correspond to dissociation probabilities found with the Pauli-Fierz Hamiltonian. 
}
\label{fig:dse} 
\end{figure}

The Pauli-Fierz model for a single vibrational mode coupled to a single field mode can be written as \cite{Flick2017,Fischer2021}
\begin{equation}\label{eq:HPF}
\hat{\mathcal{H}}_{\mathrm{PF}} = \hat H_{\rm M} + \hat{H}_{\rm C} +  \hat{H}_{\rm L} + \frac{\mathcal{E}_{0}^2}{\omega_{\mathrm{c}}}[\hat{d}(q)]^{2}, 
\end{equation}
where $\hat{H}_{\mathrm{M}}$, $\hat{H}_{\mathrm{C}}$ and $\hat{H}_{\mathrm{L}}$ are the molecular, photonic and driving Hamiltonians defined in the main text, respectively. 
The last term in Eq. (\ref{eq:HPF}) is known as the dipole self-energy (DSE) term \cite{Schafer2018,Rokaj2018}, and it depends on the vacuum field fluctuation $\mathcal{E}_{0}^{2}$ and dipole moment function $\hat{d}(q)$. 
The DSE is not included in conventional multipolar formulations of QED \cite{Power1983,Andrews2018,Feist2020,Craig1998,Buhmann2013}.

Figure \ref{fig:dse} shows photodissociation probabilities of \mbox{C-S} stretch mode of the CS$_{2}$ molecule for different values of coupling strength $\lambda_{g}$ and driving energy \mbox{$E_{\mathrm{D}}=0.055$} aJ.  
Photodissociation probabilities computed with the Pauli-Fierz model in Eq. (\ref{eq:HPF}) in the molecule-driving scenario (dashed lines) give qualitatively similar predictions as a function of time as those calculated using the Hamiltonian $\hat{\mathcal{H}}$ derived from multipolar QED formulation (solid lines). 
Hence, the results described in this work are independent of the Hamiltonian model implemented.

\end{document}